# Nd:YAG laser induced E' centers probed by in situ absorption measurements

M. Cannas[*], F. Messina

[*] Dipartimento di Scienze Fisiche ed Astronomiche dell'Università di Palermo, Istituto Nazionale per la Fisica della Materia ,Via Archirafi 36, I-90123 Palermo, Italy

## ABSTRACT

We investigated various types of commercial silica irradiated with a pulsed Nd:YAG laser radiation (4.66 eV), with exposure time ranging up to 10000 s. Transient E' centers were probed in situ by measuring the amplitude of the optical absorption band at 5.8 eV (due to E' centers) both during and after irradiation. The laser-induced absorption is observed only in natural samples, whereas the synthetic materials exhibit high toughness to radiation effect. The reported results evidence that the kinetics of E' centers is influenced by their reaction with diffusing molecular hydrogen $H_2$ made available by dimerization of radiolytic $H^0$.

* Corresponding author: M. Cannas
  Dip.to di Scienze Fisiche ed Astronomiche and INFM, via Archirafi 36, I-90123 Palermo.
  Phone: +39 09162342220, Fax: +390 916162461, e-mail: cannas@fisica.unipa.it

# 1. Introduction

E' center is one of the basic paramagnetic defects induced in high-purity $SiO_2$ glasses by ultraviolet (UV)-laser irradiation [1]. This center, apart from its electron spin resonance (ESR) features arising from an unpaired electron in a dangling orbital on a threefold coordinated Si atom ($\equiv$Si$\bullet$) [2], is also detectable by its optical absorption (OA) band peaked at 5.8 eV [3, 4]. This band is one of the main contribution to the degradation of the optical transmittance of silica glasses in several applications such as optical fibers or components in lasers. Several studies have pointed out that the generation of E' center upon laser exposure is associated with the transformation of native precursor center via different mechanisms controlled by one- or two-photons absorption [5-9]. On the other side, in a recent work based on room temperature isothermal annealing experiments, we evidenced that a fraction of photoinduced E' centers in natural silica are passivated in the post irradiation stage through diffusion-limited reactions with molecular hydrogen $H_2$ [10].

The use of spectroscopic techniques able to probe in situ the variations induced on optical and ESR signals upon UV irradiation provides an useful method to study the interplay between formation and annealing of defects. Until now, in situ luminescence measurements have been used to investigate the generation and reaction dynamics of another paramagnetic defect of fundamental interest, the non bridging oxygen hole center (NBOHC, $\equiv$SiO$\bullet$), induced in silica by $F_2$ laser photons (7.9 eV) by photolysis of SiO-H bond [11]. At variance, most of the works dealing with the generation of E' centers by UV laser photons have focused mainly on the E' center stationary concentration measured after irradiation [5, 7-9]. To our knowledge, only a few works have investigated the transient E' centers by monitoring the transmittance at 5.8 eV during irradiation [12-14]. In this paper we report experimental results obtained by in situ absorption spectra performed during and after laser exposure. Our aim is to clarify the reaction dynamics with diffusing hydrogen, which influences the kinetics of the defect.

## 2. Experimental Methods

We investigated silica samples of commercial origin (5×5×1 mm$^3$ sized) representative of four types according to Hetherington classification [15]: I) natural dry EQ906; II) natural wet Herasil 1; III) synthetic wet Suprasil 1; IV) synthetic dry Suprasil 300. UV irradiation was done at room temperature by a pulsed Nd:YAG laser (Quanta System SYL 201) working in IV harmonic mode (hν=4.66 eV) each pulse having 5 ns duration; the samples were placed with the minor surface (5×1 mm$^2$) perpendicular to the laser beam. Irradiation doses ranging from 100 up to 10000 pulses were accumulated on different samples of each silica type, at a repetition rate of 1 Hz and energy density of 40mJ/cm$^2$ per pulse. In situ OA spectra were carried out by an optical fiber spectrophotometer (AVANTES S2000) with a $D_2$ lamp source providing ~2 μW light power on the larger surface (5×5 mm$^2$) of the sample; the transmitted light was detected by a 2048 channels CCD linear array over an integration time of 3 ms. Detection timing of OA spectra was triggered at the same repetition rate (1 Hz) of laser pulses; both during irradiation, when they were collected during each inter-pulse, and in the post-irradiation stage. Stationary concentration of induced E' centers was measured by comparing the ESR signal, acquired by a spectrometer (Bruker EMX) working at 9.8 GHz, with a reference sample containing a known concentration of E' centers [4].

## 3. Results

In Fig. 1 the laser induced changes in OA spectra detected by in situ measurements in the EQ906 sample are shown. During an irradiation of 10000 pulses, Fig. 1(a), the difference between the spectra before and after increasing exposure times evidences the growth of the band (gaussian shaped) centered at 5.83±0.03 eV, full width at half maximum (FWHM) of 0.70±0.04 eV associated with the E' centers. The amplitude of the 5.8 eV band increases up to α(5.8-eV)=1.40±0.05 cm$^{-1}$, which corresponds to a concentration of ~$2.2\times10^{16}$ cm$^{-3}$ on the basis of its cross section value, $\sigma=6.4\times10^{-17}$ cm$^2$ [10]. Spectra acquired for 3600 s after the end of irradiation (Fig. 1(b)) evidence a progressive reduction of the 5.8 eV band down to 0.80±0.05 cm$^{-1}$.

Exposures to 10000 laser pulses performed on the other silica materials allow to compare the effectiveness of UV photons in inducing E' centers in the four silica types. In Fig. 2 the kinetics of the absorption coefficient $\alpha$(5.8-eV) monitored during and after irradiation are shown. In the EQ906 sample, the 4.66 eV laser radiation induces a continuous growth of the 5.8 eV band with progressively decreasing slope. We estimate the initial slope over the first 50 s, in which the curve is approximately linear, $S_0=(1.6\pm0.1)\times10^{-3}$ cm$^{-1}$ s$^{-1}$. When the laser source is switched off, we observe a sudden change of slope due to the begin of the isothermal annealing of E' centers, already active in a timescale of a few seconds. On increasing the delay after the laser irradiation, $\alpha$(5.8-eV) partially reduces, its variation being almost complete after 3600 s. For the Herasil 1 sample the laser irradiation induces the growth of the absorption at 5.8-eV, the initial slope being $S_0=(1.1\pm0.1)\times10^{-3}$ cm$^{-1}$ s$^{-1}$, which tends to saturate with time to a constant value of (0.65±0.05) cm$^{-1}$. In the post irradiation stage, we observe the reduction of $\alpha$(5.8-eV) that tends to less than 10% of maximum value after a few hours. As regards the two synthetic materials, we do not observe any increase of the 5.8 eV band during irradiation, at least within our experimental uncertainty of 0.05 cm$^{-1}$.

To better evidence the role of post-irradiation annealing in governing the generation of E' centers, in Fig. 3 we show the stationary concentration measured EQ906 samples a few days after the end of irradiations with increasing number of pulses. These data evidence that E' centers accumulate more and more with dose, following a sub-linear growth that differs from the transient concentration measured during laser exposure. Hence, the post-irradiation reduction of E' centers, measured as the difference $\Delta$[E'] between transient and stationary values, depends on dose as shown in the inset.

## 4. Discussion

The above reported results, based on in situ OA measurements, evidence that E' centers are induced by UV rays in natural silica. On the basis of the initial slope $S_0$ of the growth kinetics, we estimate the parameter $Q=(S_0/\sigma)\times(h\nu/I)$, where $S_0/\sigma$ is the initial E' centers concentration growth rate, I is the intensity of laser beam and $h\nu =4.66$ eV is the photon energy. Q measures the extinction coefficient of laser photons due to the generation of E' centers in silica materials and amounts $\sim 5\times 10^{-4}$ cm$^{-1}$ for EQ906 and $\sim 3\times 10^{-4}$ cm$^{-1}$ for Herasil 1. On the other side, the absence of the OA at 5.8 eV in the spectra measured for Suprasil 1 and Suprasil 300 samples implies that E' centers are not induced, at least within a concentration of $\sim 7\times 10^{14}$ cm$^{-3}$. The higher toughness of synthetic materials to radiation damage suggests that the generation of E' centers upon UV ray exposure is associated with the conversion of precursor defects peculiar to manufacturing procedure of natural silica. The defects most commonly regarded as E' precursors are the oxygen deficient centers (ODC) and the Si-H and Si-Cl bonds [7]. However, we acknowledge that our data do not allow to clarify unambiguously the conversion mechanism, though we note that 4.66 eV Nd:YAG laser radiation really excites a OA band related to ODC [16], whereas the occurrence of UV transitions of Si-H and Si-Cl has not been yet clarified [1].

Now we discuss the influence of the annealing processes in governing the transient behavior of E' centers. A first point emerges from the observation that the post-irradiation decay of E' centers mainly occurs in a timescale of $(10^3\text{-}10^4)$ s, which is comparable with the irradiation time of our experiments (Fig. 2). Then, the processes responsible for the annealing are active also during UV exposure and compete with the laser-induced generation. In this view, we stress that the saturation reached during irradiation could be caused by the equilibrium between generation and annealing rather than the exhaustion of precursor defects.

Furthermore, the post-irradiation decay plays a main role in determining the stationary concentration of E' centers accumulating in silica. Our results evidence a difference between dry and wet natural silica when the annealing kinetics is complete: (i) in dry samples, we measure residual E' centers whose concentration increases with dose; (ii) in wet samples, the E' centers induced during irradiation are almost completely passivated in the post irradiation stage.

The origin of post-irradiation decay is usually ascribed to the reaction of E' center with molecular hydrogen diffusing at room temperature in silica matrix :

$$\equiv Si \bullet + H_2 \rightarrow \equiv Si\text{-}H + H^0 \qquad (1)$$

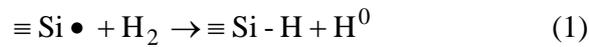

where the released $H^0$ can be involved in further reactions with E' centers, and the kinetics of reaction (1) is governed by the $H_2$ mobility [17]. The validity of this model has been recently proved by isothermal annealing OA and ESR experiments performed on UV-irradiated natural silica, in which one observes that the E' center passivation is the main channel reducing the hydrogen content [10, 18]. Then, the total post-irradiation decrease $\Delta[E']$ equals one half of the amount of $H_2$ available at the end of irradiation $[H_2]_0$. In this framework, inset of Fig. 3 shows that $[H_2]_0$ in dry samples increases with dose up to ~5000 pulses so evidencing the radiolytic origin of the hydrogen, which can be made available in atomic form $H^0$ via photoinduced breaking of SiO-H or Si-H bonds and then dimerize to form $H_2$ [6,8,17]. This result allows us to shed light on the difference between dry and wet samples, outlined in the points (i) and (ii), that we quantify by the ratio $[E']_0/ 2[H_2]_0$. In the dry silica, this ratio is >1, so meaning that E' centers are produced in higher concentration than hydrogen atoms. In wet silica, since $[E']_0/ 2[H_2]_0 \approx 1$, E' centers and hydrogen are produced approximately in the same amount. This is consistent with a model in which E' centers and hydrogen in natural wet $SiO_2$ are generated from the same precursor Si-H, as also proposed in previous works [6].

## 5. Conclusions

Our data show that Nd:YAG laser irradiation induces in natural silica the absorption band at 5.8 eV related to E' centers and hydrogen, the latter made free by rupture of preexisting bonds, such as Si-H and O-H. The reaction between E' centers and hydrogen influences the transient concentration of the defects both during irradiation and in the post-irradiation stage. These results prove the usefulness of in situ spectroscopic techniques to perform comprehensive studies on conversion processes leading to the photogeneration of E' centers and to their isothermal annealing governed by diffusing molecular hydrogen.

### Acknowledgements

The authors wish to thank Prof. R. Boscaino and his research group at University of Palermo for their support and enlightening discussions. Technical assistance of G. Lapis and G. Napoli is also acknowledged. This work is part of a national project (PRIN2002) supported by the Italian Ministry of University Research and Technology.

# Figure captions

**Figure 1**: Absorption difference spectra before and during irradiation with 10000 laser pulses (a); after various post irradiation delay times up to 3600 s (b).

**Figure 2**: Kinetics of the absorption coefficient at 5.8 eV recorded during and after 10000 laser pulses irradiation in 4 silica types: I) EQ906; II) Herasil 1; III) Suprasil 1; IV) Suprasil 300.

**Figure 3**: Concentration of E' centers induced during laser irradiation (full symbol), data from Fig. 2, and measured at the end of the annealing time (open symbol) as a function of accumulated laser pulses in EQ906 silica samples. The inset shows the reduction of E' centers.

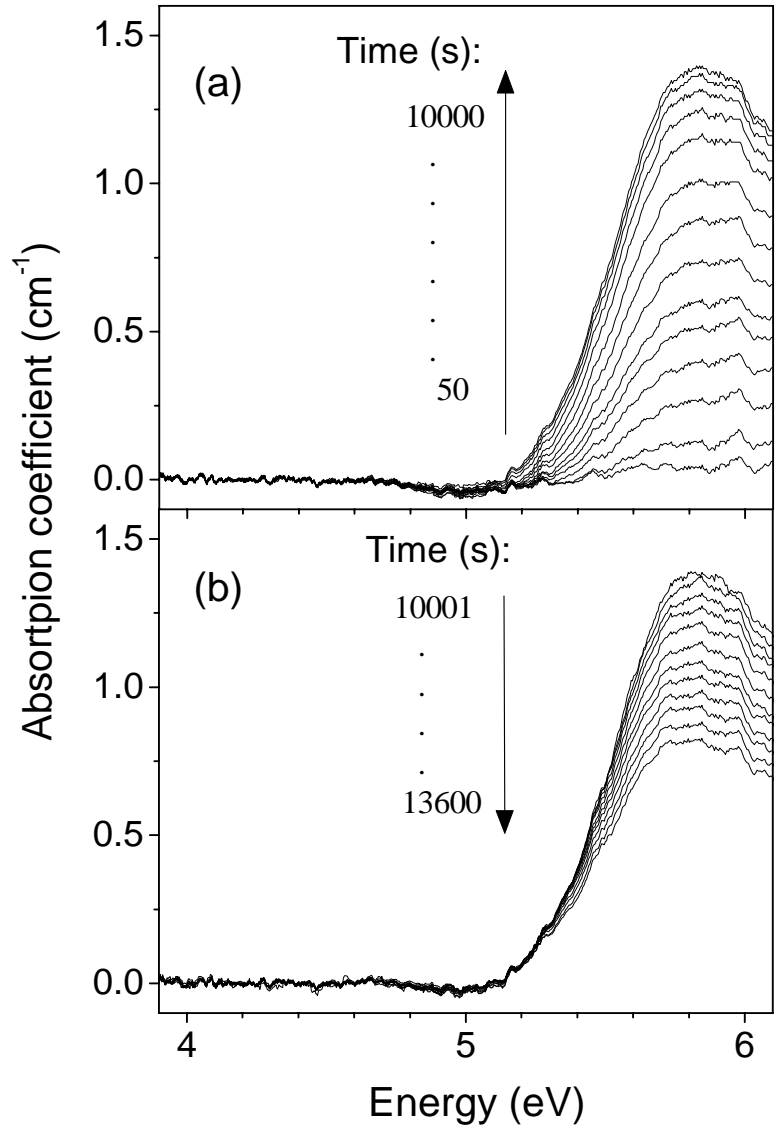

Figure 1

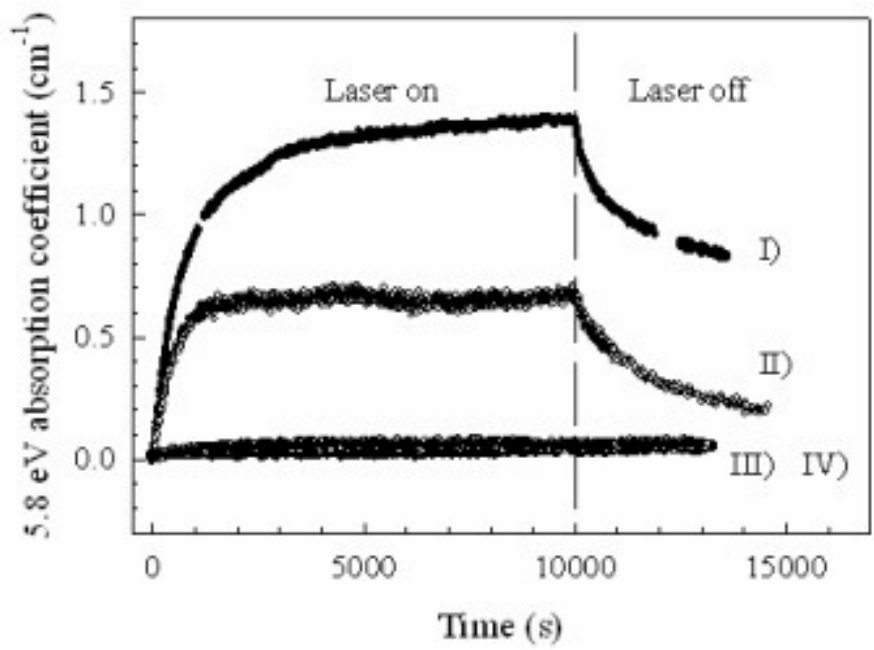

Figure 2

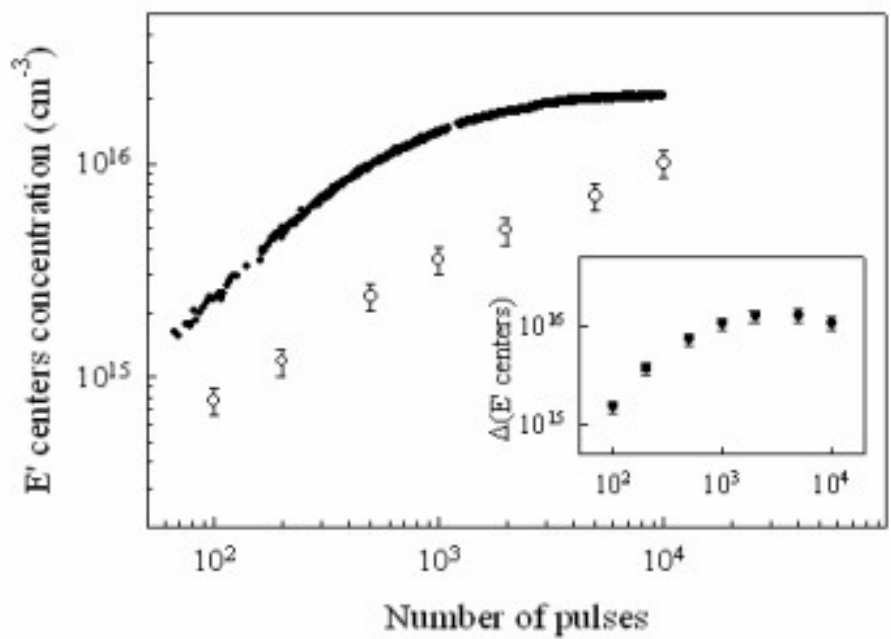

Figure 3